# THE BISCARI ARCHIVE. A CASE STUDY OF THE APPLICATION OF THE TRANSKRIBUS TOOL

Salvatore Spina[1]

[1]University of Catania, Italy – salvatore.spina@unict.it

...

**Abstract**

The 'Paternò-Castello Principi di Biscari' Archive, preserved at the State Archives of Catania – amongst one of the most crucial family archives – is, in the light of a digital historical methodology, the best computable historical heritage for demonstrating the applicability of applying an HTR tool, such as Transkribus, to digitised historical documents.

**keywords**

Transkribus, Filemaker, Gephi, Princes Ignazio, Princess Anna, Michele Maria Paternò.

**INTRODUCTION**

The archive of the Paternò-Castello family, princes of Biscari, kept at the State Archives of Catania, represents a valuable set of computable historical data. Moreover, its heterogeneity may support the hypothesis that it is possible to transform the archival heritage of History into Big Data. This means that all kinds of historical information can be digitised. Therefore, a proper digitisation process would help create a historical documentary apparatus capable of assisting the historian in his efforts to reconstruct past events. Due to its characteristics, precisely because it is the cultural, social, and political expression of a dynamic community, the Biscari Archive stores documents with unique peculiarities that a historian cannot disregard. Applying sophisticated computer tools is the only way to bring out all the data from the sources. Indeed, human physiological limitations do not allow for the analysis of large amounts of data and information. Fortunately, the development of artificial intelligence, deep learning tools and neural networks can compensate for this inherent difficulty in human nature.

However, some clarifications are in order. First and foremost, historical documentation of the Modern Era is composed mainly of manuscripts, representing an obstacle, as they are, to the possibility of using computer tools.

Archival documentation is mainly composed of manuscripts, which is a significant obstacle to the possibility of using computer tools for a study concerning an event that took place in the Modern Age.

Secondly, a lot has already been said about digitisation and digitalisation. However, the two terms reproduce a conception in most scholars: "digitisation" is like "taking a picture". According to "analogue" historians, this is one of the assumptions on which to build the Digital History Statute'. Most historians believe that the digital methodology consists of scanning or photographing some archival document and uploading it to a database. Indeed, one of the initial stages of any historical study is the retrieval and organisation of the. However, in a digital dimension that digitises objects, places and people, a photo cannot be read by Artificial Intelligence in the same way the historian reads the text of the image. For a computer, a photo is a set of data, such as the colour code – *e.g.* #0000 is 'black', #654321 is 'brown', #FF0000 is 'red', and so on –, the definition, the number of pixels, the place where it was taken – if the camera is equipped with GPS – and other similar information[1].

Nevertheless, if we consider the text of any manuscript, no computer can figure out all the information it carries because it is inaccessible even to the Turing machine. If this were the case, the computer could only reproduce words without relation to reality because language is a code structured on non-explicit elements, inconsistency between word and action and other aspects that computers cannot process. Furthermore, even if a computer can analyse a digital text, it cannot read and recognise a handwritten word, not even a single consonant or vowel.

---

[1] S. Spina, *Digital History. Metodologie informatiche per la ricerca storica*, Edizioni Scientifiche Italiane, Napoli 2022.

Still, this assumption brings us to some key points: is photographing the right way to digitise an archival heritage? Can a computer read handwritten text? How could digitisation and computer tools change the craft of historians?

A study on historical sources of the Biscari Archive – Catania, Italy[2] – can help us solve the dilemma and demonstrate the feasibility of using HTR technology at various stages of historical research.

Within this archive are 2,000 folders with hundreds of thousands of sheets documenting legal disputes, political decisions, and business and personal letters. The 'Correspondence' section consists of some 42,493 papers, in which, in folder 1642, we find 591 sheets, constituting 366 letters and a manuscript by Emile Rousseau.

Although a 'quantitative' approach would have been more appropriate to choose the initially computable documentary material – such as an "income and expenditure register" – a folder displaying non-numerical information was opted for. This is because the creation of an innovative model of historical research, as demonstrated by Ruth and Sebastian Ahnert and Kim Albrecht (https//tudor-networks.net), requires the development of technological systems capable of analysing heterogeneous data, such as those chosen in our case.

The documents acquisition was performed with the Nikon D610 equipped with the AF-S Nikkor 24-120mm f/4G ED VR lens, and the photos taken were collected in a database created with Claris Filemaker 19, a software that was chosen since a website is not planned to be built at this stage of the project.

After the metadata[3] insertion phase, the photos were converted to PDF format to reconstruct the 366 letters, and 53 of these were uploaded to the Transkribus server to obtain the automatic transcription.

## 2. TRANSKRIBUS & BISCARI ARCHIVE

Due to its Java technology, Transkribus[4] (https://readcoop.eu/transkribus/?sc=Transkribus) allows the creation of workflows based on "deep neural networks"[5], which can be trained to recognize a particular handwriting. One of the options required to get the best result is that the texts are written by the same hand.

For this reason, 53 letters were chosen, 28 letters from Michele Maria Paternò – Prior of Messina – to Princess Anna Maria Morso Bonanno – wife of Ignazio Paternò, fifth prince of Biscari –, and 25 letters from Marquis Giovanni Fogliani Sforza D'Aragona, Viceroy of Sicilia, sent to different addresses.

The 53 letters were selected after analysing the network of people (fig. 1) who corresponded with the Prince of Biscari or other members of the house of Paternò-Castello.

---

[2] G. Calabrese, *L'archivio della famiglia Paternò Castello principi di Biscari: inventario*, Os. n.!, Catania 2003.

[3] Sender, consignee, date and place of issue, place of destination, named persons and places, *regesto* and keywords.

[4] P. Kahle *et al.*, *Transkribus. A Service Platform for Transcription, Recognition and Retrieval of Historical Documents*, *2017 14th IAPR International Conference on Document Analysis and Recognition (ICDAR)*, 2017, v. 04, pp. 19–24; N. Milioni, *Automatic Transcription of Historical Documents. Transkribus as a Tool for Libraries, Archives and Scholars*, 2020.

[5] H. Aghdam, E. Heravi, *Guide to Convolutional Neural Networks. A Practical Application to Traffic-Sign Detection and Classification*, 2017; G. Amato *et al.*, *Visual Recognition of Ancient Inscriptions Using Convolutional Neural Network and Fisher Vector*, in «Journal on Computing Cultural Heritage», 9, 4 2016, pp. 1–24; M. Boillet *et al.*, *Multiple Document Datasets Pre-training Improves Text Line Detection With Deep Neural Networks*, in «2020 25th International Conference on Pattern Recognition (ICPR)», 2021, pp. 2134–2141.

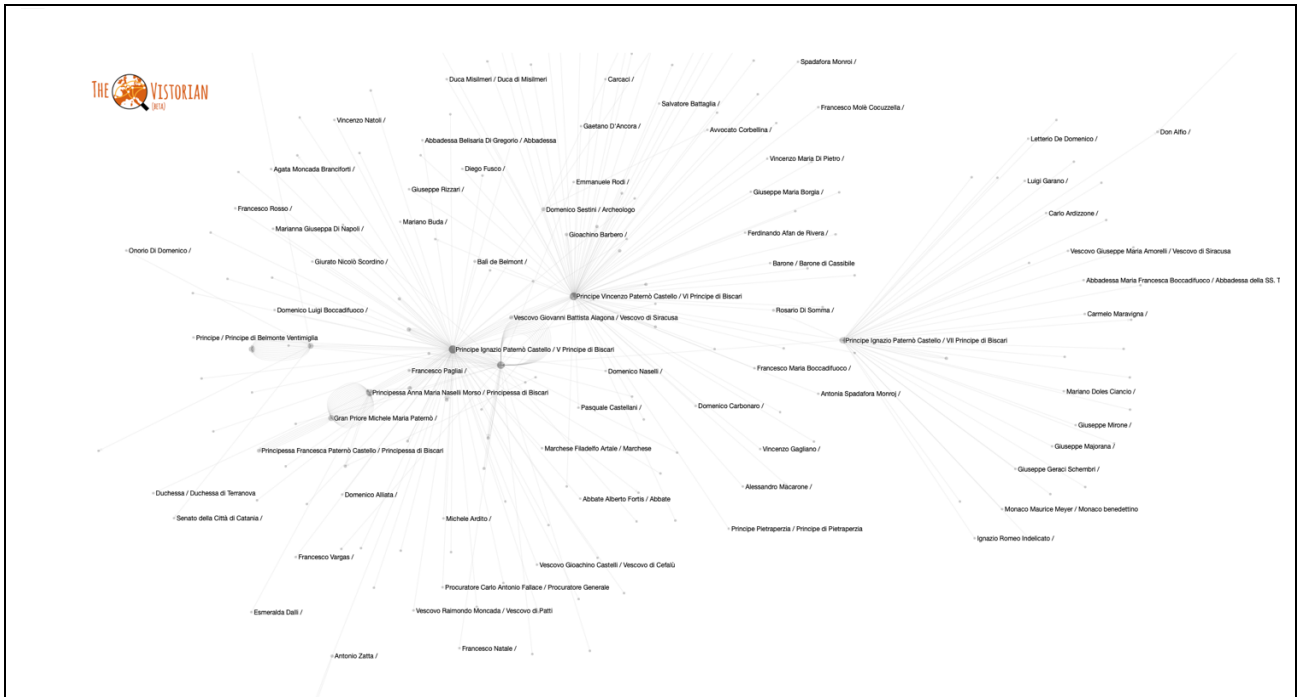

*Figure 1. Sender and consignee.*

The database configuration allowed us to index each person and place the 367 records into a relational structure. Thanks to the geo-coordinates and the application of 'The Vistorian', 'Palladio', and 'Gephi' tools, it was possible to (1) place the various senders in territories (*e.g.* Sicily, Italy, Europe – fig. 2); (2) determine the different relations between the data (*e.g.* sender-consignee).

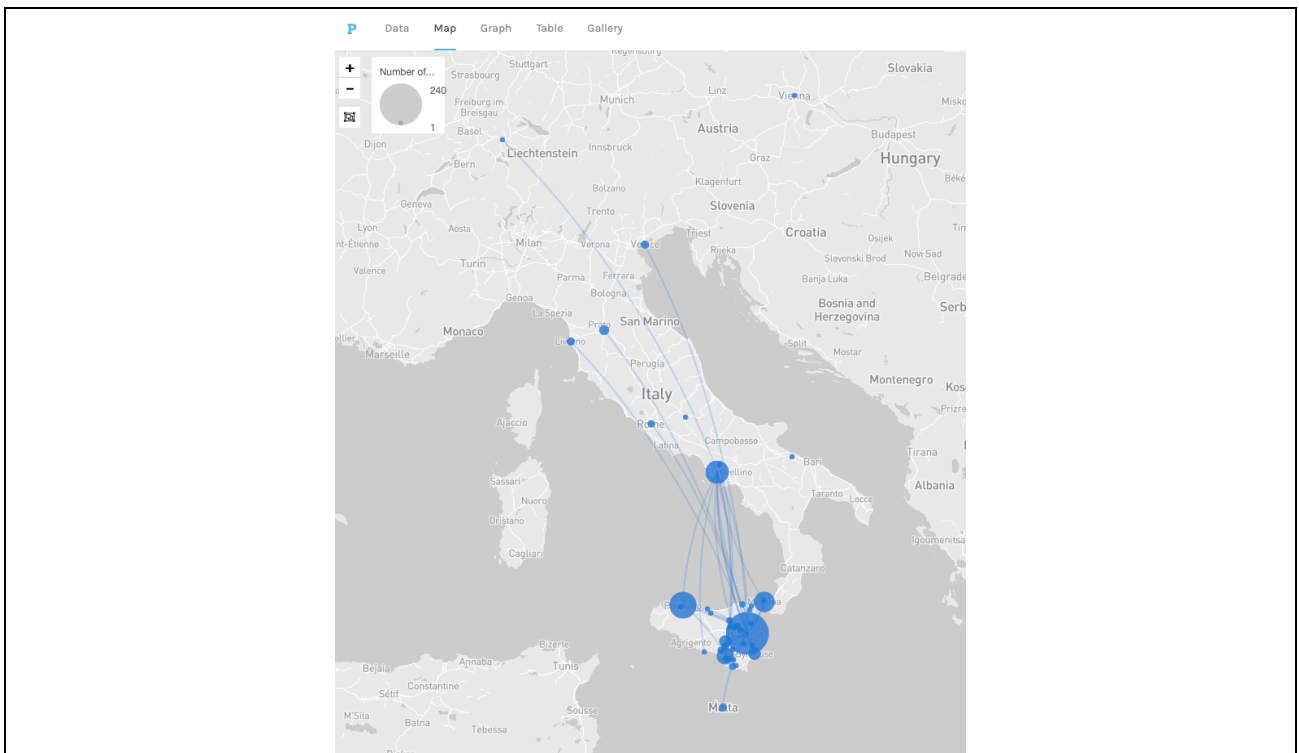

*Figure 2. The network of the place.*

However, a crucial question persists: how does a historian identify a group of letters written by the same hand? One of the fundamental aspects of Digital History is its ability to narrate by "visualising". Through an editorial design product, *i.e.*, it is possible to display information which allows a scholar to be guided towards a comprehensive view of the object of study. Moreover, among the most helpful and widely used approaches, Historical Network Analysis (HNA) decisively meets historians' need to make sense of information.

The HNA postulates that men build relationships with fellow men and the surrounding space – even in the absence of other subjects. Wherever there is a man, there is a relation. Where there is a 'node' (subject), there is an 'arc' (edge; the relationship). This schema functions as a human paradigm (*i.e.* "the network framework shapes the way we interpret the world around us") that becomes the foundation of the Network Analysis approach: everything can be reconstituted visually into an open schema – that can be continuously filled in – that provides different information depending on the object of research.

Thanks to Gephi, it was possible to identify an appropriate number of letters written by the same sender (fig. 3) – thus by the same hand – ensuring the uniformity required by the Transkribus tool.

Within the Biscari Archive, precisely at folder 1642, section 'Correspondence', can be found 28 letters written by Michele Maria Paternò, Prior of Messina. He was a religious man very close to the Biscari family, especially to Princess Anna, the addressee of this correspondence. While the other 25 letters chosen for the test were written by Marquis Fogliani.

Both groups of letters (a total of 112 sheets) were sorted and merged into two separate pdf files, which were subsequently uploaded to the Transkribus server. The small size of the corpus certainly facilitated rapid analysis and automatic transcription, even though the tool's purpose is to allow the transcription of larger archival material. After all, Transkribus is meant to meet one of the benchmarks of historical research methodology in the digital environment: automatically transcribing many handwritten texts that historians would never be able to record during their research.

Here, the number of letters is sufficient to understand whether Traskribus is able to transcribe them correctly and, above all, whether the HTR model for the Italian language has been adequately trained – despite the inevitable reluctance of the Italian scientific community to use state-of-the-art IT tools. Whilst the international scientific community aspires to build an adequately digitised historical heritage, the Italian's is failing to break out of the constraints of the legal protection of archival heritage, leading to a delay in digitising historical heritage. In addition, Italy lacks a scientific mindset to ensure the dissemination of its national archival heritage, unlike, for example, Finland, where, thanks to the National Archives project, the Central Government wanted to restore to its community the archival heritage related to the Second World War, the judicial registers and property inventories of the Finnish nobility. To do this, it involved students and scholars in training a model to obtain transcriptions with minimal margins of error. Detailed and in-depth training will enhance artificial intelligence, enabling Transkribus to automatically transcribe the entire human heritage in all languages and eras of the past.

A further core feature of Transkribus is that digital transcription projects of archival heritages can be open to the entire community, who thus participate in the training phases, thus realising the grand goal of Public History.

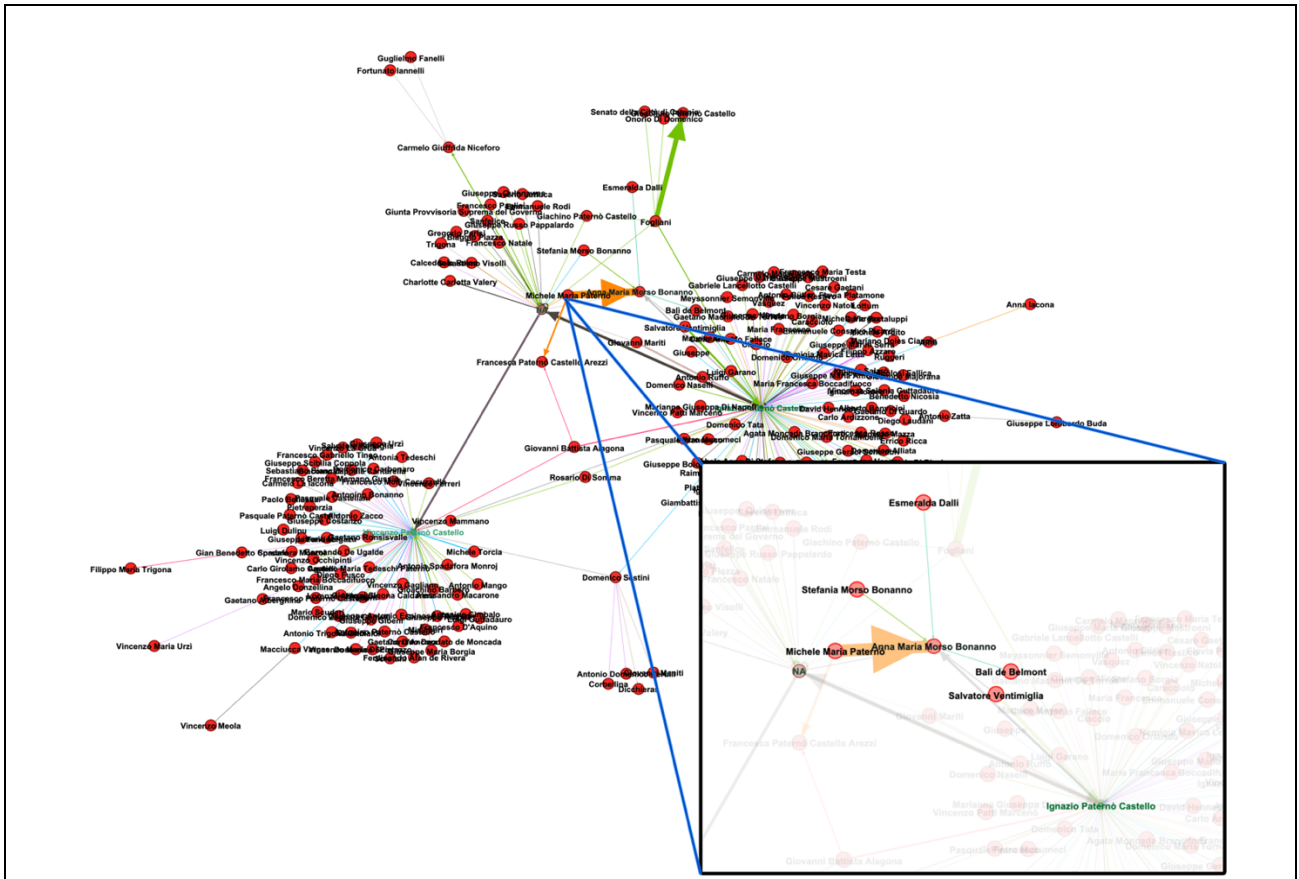
*Figure 3. Correspondence between Michele Maria Paternò and Anna Maria Morso Bonanno.*

### 3. OUTCOME
The automatic transcription of the two sets of letters took place without any training activity because the primary intent was to understand the accuracy achieved by the public Transkribus AI model.
For transcription, we applied the "Italian Administrative Hands 1550-1700" model created by Jake Dyble, Antonio Iodice, Sara Mansutti and Rachel Midura, whose CER (Character Error Rate) is 9.15 per cent. This rate represents an excellent result for a public model in the automatic transcription of manuscripts kept in Italian archives. The latter, in fact, hold the most heterogeneous historical heritage in Europe, not only because of the different writing styles but also because of the presence of different languages and institutions – *e.g.*, Latin in the Vatican, Spanish in the Kingdom of Sicily and the Duchy of Milan, Italian in the Grand Duchy of Tuscany.
The model is applied and transcribed when the two PDF files are uploaded to the Transkribus server. In the case of Marquis Fogliani's letters, the error rate is 7% (*e.g.*, in the first letter, 6 out of 82 words are not transcribed correctly), and the interval remains constant for almost all sheets.
The error interval is much higher for the 28 letters written by Michele Maria Paternò. In fact, it exceeds 10% due to the presence of incorrectly written Italian words and the presence of words unknown to the Transkribus model. Also, in the second letter, it can be seen that the graphemes "V.S." and "S.M," although present in the segmentation of the text image, are not transcribed. This condition results in a higher error rate because the tool not only did not transcribe even incorrectly but did not allow any correction by the user. The latter cannot identify the word in the window intended for digital transcription and its correction.
Given the numerous transcription errors, it was decided to train a model based on the "Italian Administrative" model, starting with the 24-page "ground truth" (pages 1-5, 7-9, 11-18, 20-24, 26, 27, 35). Once completing the training and forming the model – "Michele Maria Paternò_Archivio

Biscari_Model" –, the CER of the Training Set dropped to 0.68% and that of the Validation Set to 9.42% (fig. 4).

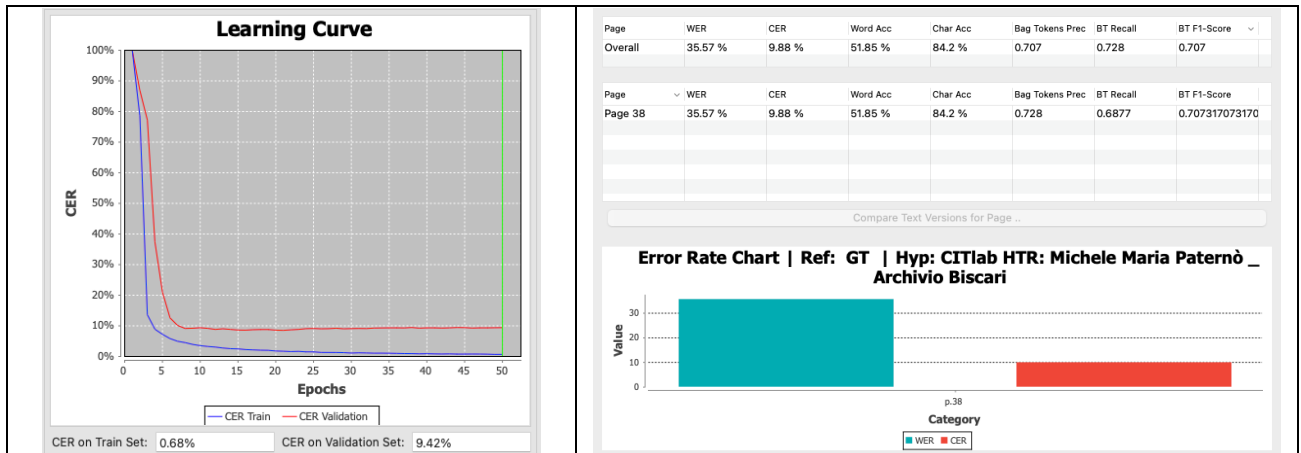
*Figure 4. Correspondence between Michele Maria Paternò and Anna Maria Morso Bonanno.*

From the comparison (*e.g.*, page 38 - see Figure 5) between the "ground truth "(manual) transcription and the automatic one, on the model just mentioned, the CER stands at 9.88%, yet that of the validation set returns a relatively high percentage (35.57%), which is indicative of training that needs to be carried out on a more substantial ground truth (50 pages, at least) to provide a quantitatively and qualitatively accurate transcription. However, as can be seen (Figure 7), the application of the "Michele Maria" model resulted in a transcript whose CER is 8.07%/26.29 and with minor errors, which do not compromise the scholars' inferring but make the computational analysis unreliable.

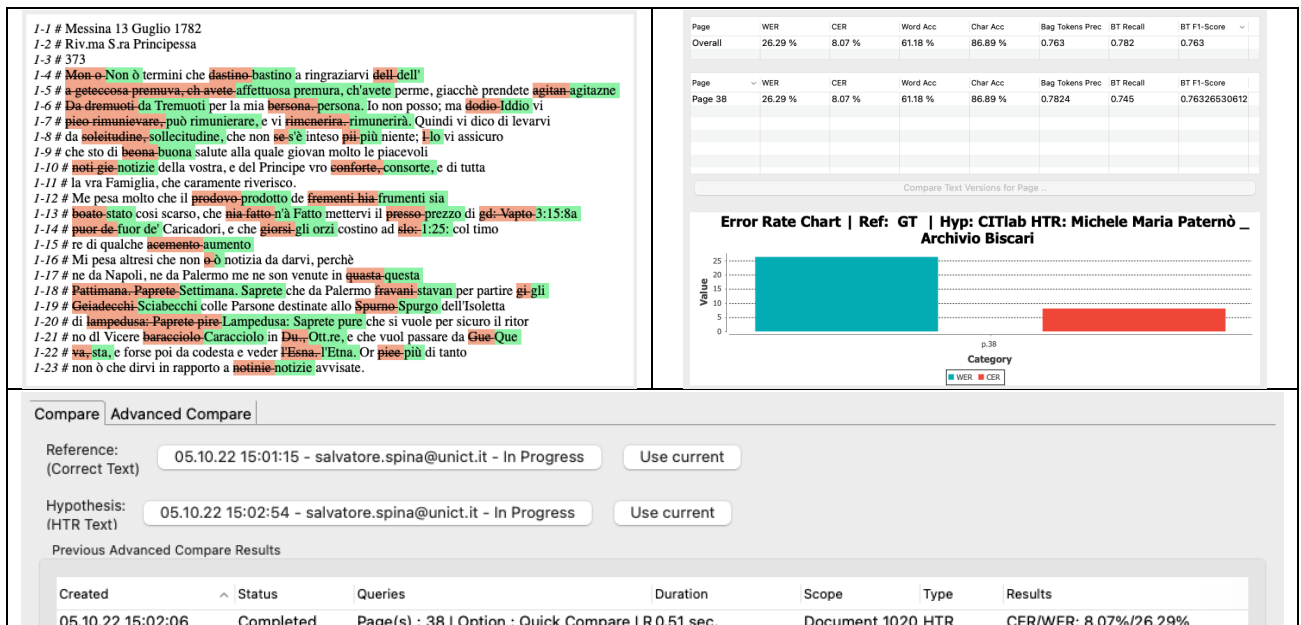
*Figure 5. Comparison set.*

## 4. TRANSKRIBUS E FILEMAKER

(1) A research project in History starts with the organisation of primary sources. (2) "Analogue" historians have to transcribe the sources, analyse them in detail and try to seize all the meanings through their intuitive and intellectual faculties alone. (3) "Digital" historians use digital archives and databases, in which he arranges sources and the information they have drawn from them. (4) Each stage can take advantage of computer technology, allowing historians to process more information and coordinate many more sources.

The study conducted at Biscari Archive documents aims to show how these technologies can dialogue with each other and enable historians to reconstruct past events thanks to more articulate and detailed information.

Transkribus allows transcriptions to be exported in PDF files, which can be analysed, processed, and organised in databases. This enables searches through an interface and a relational system highlighting the novel information.

Thus, Transkribus makes it one of the essential tools for building the Big Data of History. Moreover, it responds to Gardin's assumption that the creation of databases and ontologies cannot be separated from the correct 'encoding and formalisation' phase, which enables the scholar to codify historical texts into a computable version.

To this aim, the 53 transcriptions from the Biscari archive were entered into a database created with Filemaker 19 (fig. 6), a software that enables the creation of a relational structure between the data within it.

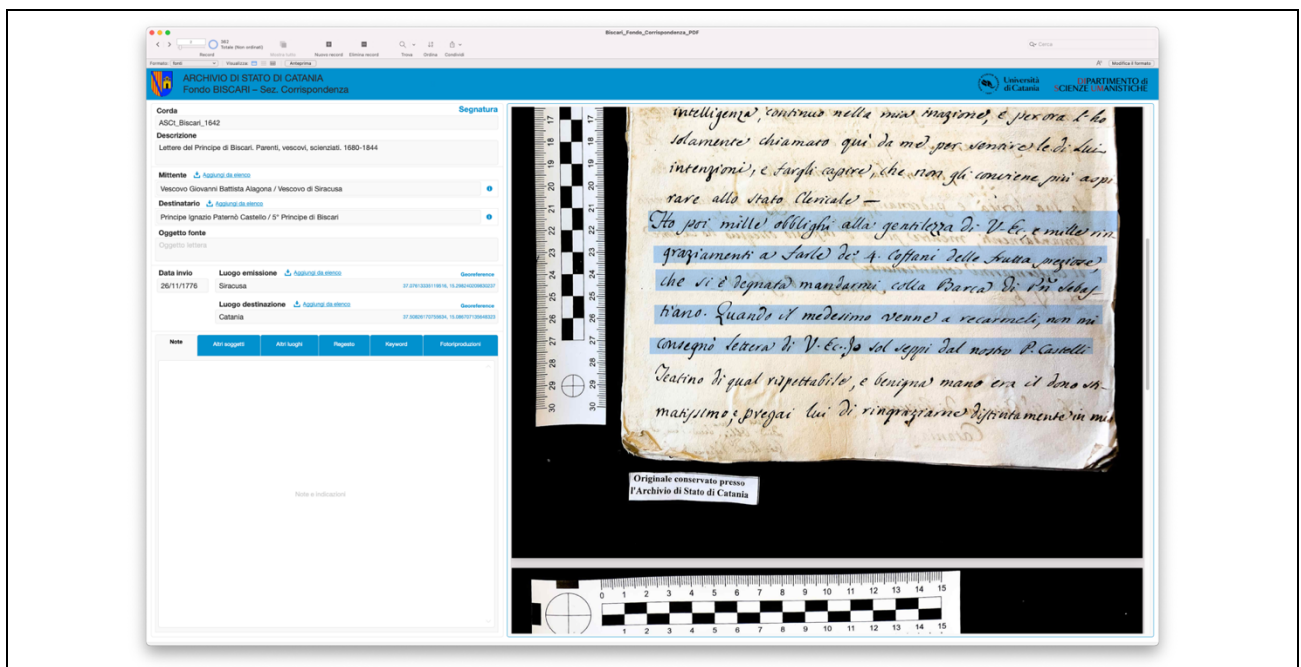

*Figure 6. The Database, Transkribus and Filemaker.*

For its part, Filemaker, with the update to version 19.5, has also implemented a new JSON function, which allows for the differentiation of numbers and text and the use of system libraries, OCR and HTR images inserted in a container field – a possibility that not only reflects the need to deal with the digitisation of photographic sources but also makes the databases ready for the creation of Machine-Readable sources.

Transkribus aims to demonstrate the need for upgrading historians' skills in a digital environment, focusing on the most sensitive and essential aspects of source analysis: organisation and transcription.

## 5. ACHIEVEMENT

The term 'digitisation' has pervasively entered the vocabulary of our society. This concept, however, is very often misused. Firstly, three concepts are lumped together (digitisation, digitalisation, and digital transformation). Secondly, for "analogue" humanists, this term is synonymous with "facsimile", *i.e.* an image to be observed on a monitor, without considering that the digitisation process is quite different from the simple act of photographing/scanning an archival document.

Every document is an aggregate network compound of pieces of information, and only its proper digitalisation will allow us to create digital texts that can enrich our knowledge of past events, thanks to tools such as Network Analysis, which allows the intuitive processes of the historian to reach a level previously unattainable. Networks of people, places and events become, in fact, the necessary background for a historical ontology, but the necessary step remains, irrefutably, the automation of the transcription of manuscript sources on which historical research is based. The information stored in a photographed text escapes the computer. To the latter, images are just meaningless sets of pixels. It is the semantic context of a source, the meaning of words, that makes a document instrumental to reconstructing a historical event.

Thus, the Biscari Archive is an essential asset to reconstructing the History of Catania and of many events that marked the History of Sicily. However, every piece of information contained therein is yet to be identified and analysed. For this reason, our trial is intended as a call to look at the archival heritage in the light of Digital History and ICT paradigms. Historical research requires an essential investment in digitising archival sources, considering that the digital world requires a specific configuration (through Transkribus and other tools) that goes beyond simple photographic acquisition.

**REFERENCE**
H. Aghdam, E. Heravi, *Guide to Convolutional Neural Networks. A Practical Application to Traffic-Sign Detection and Classification*, 2017
G. Amato, F. Falchi, L. Vadicamo, *Visual Recognition of Ancient Inscriptions Using Convolutional Neural Network and Fisher Vector*, in «Journal on Computing Cultural Heritage», fasc. 9, 4 2016, pp. 1–24
M. Boillet, C. Kermorvant, T. Paquet, *Multiple Document Datasets Pre-training Improves Text Line Detection With Deep Neural Networks*, in «2020 25th International Conference on Pattern Recognition (ICPR)», 2021, pp. 2134–2141
G. Calabrese, *L'archivio della famiglia Paternò Castello principi di Biscari: inventario*, Os. n.!, Catania 2003
M. C. Calabrese, *Il Gran Priore dell'Ordine di Malta e il suo servo. Un episodio di schiavitù mediterranea nel XVIII secolo*, in «Nuova rivista storica», fasc. 100, 3 2016, pp. 907–976
F. D'Avenia, *Le commende gerosolimitane nella Sicilia moderna: un modello di gestione decentrata*, 2000, pp. 35–86
P. Kahle, S. Colutto, G. Hackl, G. Mühlberger, *Transkribus. A Service Platform for Transcription, Recognition and Retrieval of Historical Documents*, 2017 14th IAPR International Conference on Document Analysis and Recognition (ICDAR), 2017vol. 04, , pp. 19–24
N. Milioni, *Automatic Transcription of Historical Documents. Transkribus as a Tool for Libraries, Archives and Scholars*, 2020
S. Spina, *Digital History. Metodologie informatiche per la ricerca storica*, Edizioni Scientifiche Italiane, Napoli 2022